\documentclass[sigconf]{acmart}

\usepackage{multirow}
\usepackage{pifont}
\usepackage{makecell}
\usepackage{enumitem}
\usepackage{bm}

\copyrightyear{2026}
\acmYear{2026}
\setcopyright{cc}
\setcctype{by}
\acmConference[WWW '26]{Proceedings of the ACM Web Conference 2026}{April 13--17, 2026}{Dubai, United Arab Emirates}
\acmBooktitle{Proceedings of the ACM Web Conference 2026 (WWW '26), April 13--17, 2026, Dubai, United Arab Emirates}
\acmPrice{}
\acmDOI{10.1145/3774904.3792716}
\acmISBN{979-8-4007-2307-0/2026/04}

\begin{document}

\title{Hyena Operator for Fast Sequential Recommendation}

\author{Jiahao Liu}
\affiliation{%
  \institution{Wuhan University of Technology}
  \city{Wuhan}
  \country{China}}
\email{jh.liu@whut.edu.cn}

\author{Lin Li}
\authornote{Corresponding author.}
\affiliation{%
  \institution{Wuhan University of Technology}
  \city{Wuhan}
  \country{China}}
\email{cathylilin@whut.edu.cn}

\author{Zhiyuan Li}
\affiliation{%
  \institution{Wuhan University of Technology}
  \city{Wuhan}
  \country{China}}
\email{zhiyuanli@whut.edu.cn}

\author{Kaixi Hu}
\affiliation{%
  \institution{Wuhan Textile University}
  \city{Wuhan}
  \country{China}}
\email{kxhu@wtu.edu.cn}

\author{Kaize Shi}
\affiliation{%
  \institution{University of Southern Queensland}
  \city{Toowoomba}
  \country{Australia}}
\email{Kaize.Shi@unisq.edu.au}

\author{Jingling Yuan}
\affiliation{%
    \department{Hubei Key Laboratory of Transportation Internet of Things}
  \institution{Wuhan University of Technology}
  \city{Wuhan}
  \country{China}}
\email{yjl@whut.edu.cn}

\renewcommand{\shortauthors}{Jiahao Liu et al.}

\begin{abstract}
Sequential recommendation models, particularly those based on attention, achieve strong accuracy but incur quadratic complexity, making long user histories prohibitively expensive. Sub-quadratic operators such as Hyena provide efficient alternatives in language modeling, but their potential in recommendation remains underexplored. We argue that Hyena faces challenges in recommendation due to limited representation capacity on sparse, long user sequences. To address these challenges, we propose HyenaRec, a novel sequential recommender that integrates polynomial-based kernel parameterization with gated convolutions. Specifically, we design convolutional kernels using Legendre orthogonal polynomials, which provides a smooth and compact basis for modeling long-term temporal dependencies. A complementary gating mechanism captures fine-grained short-term behavioral bursts, yielding a hybrid architecture that balances global temporal evolution with localized user interests under sparse feedback. This construction enhances expressiveness while scaling linearly with sequence length. Extensive experiments on multiple real-world datasets demonstrate that HyenaRec consistently outperforms Attention-, Recurrent-, and other baselines in ranking accuracy. Moreover, it trains significantly faster (up to 6× speedup), with particularly pronounced advantages on long-sequence scenarios where efficiency is maintained without sacrificing accuracy. These results highlight polynomial-based kernel parameterization as a principled and scalable alternative to attention for sequential recommendation.
\end{abstract}




\maketitle

\section{Introduction}

In modern digital platforms (e.g., Amazon\cite{he2016ups, mcauley2015image}, Steam\cite{kang2018self}), users frequently generate multiple interaction records, and the efficient, accurate prediction of users’ next actions from their historical behaviors is a core problem for improving recommendation quality and user experience. As user histories continue to grow, recommender systems must address the challenge of modeling increasingly long interaction sequences while maintaining fast inference and training efficiency—making fast sequential recommendation a critical research direction. To address this task, recent research and industrial practice have proposed a variety of methods: Markov chain-based\cite{he2016fusing, rendle2010factorizing} approaches model short-term patterns via local transitions. Sequential dependencies have been modeled by RNN-based architectures\cite{hidasi2018recurrent, hidasi2015session, li2017neural, yue2024linear}, while CNN-based architectures\cite{tang2018personalized, yan2019cosrec} capture local patterns through convolutional operations. Attention-based methods\cite{he2021locker, kang2018self, sun2019bert4rec, tian2024eulerformer, DBLP:conf/cikm/HuL0LT21, DBLP:journals/toit/HuLLS21} flexibly model arbitrary long-range dependencies through self-attention; and more recently, generative\cite{zhai2024actions} approaches attempt to handle long sequences with lower computational complexity by capturing the underlying semantics of user interactions.

Despite progress in various scenarios, each approach has inherent limitations. Markov chain and shallow statistical methods\cite{he2016fusing, rendle2010factorizing} capture local transitions. RNNs\cite{hidasi2018recurrent, hidasi2015session, li2017neural, yue2024linear} suffer from gradient propagation and efficiency issues on long sequences. Traditional convolutions\cite{tang2018personalized, yan2019cosrec}, though efficient, are limited in their receptive field scalability. Attention-based methods\cite{he2021locker, kang2018self, sun2019bert4rec, tian2024eulerformer} can model global dependencies, but they incur high training costs and substantial time and memory overheads during incremental inference. Generative\cite{zhai2024actions}  approaches offer scalability, yet they remain insufficient for jointly modeling multi-scale behaviors, including slowly evolving long-term interests and frequently occurring short-term bursts. Overall, the core challenge in sequential recommendation, particularly for long user sequences, lies in efficiently and stably capturing user behavior patterns under sparse feedback, as conventional operators often have limited representation capacity in such scenarios. Table~\ref{tab:comparison_main} summarizes the characteristics of representative sequential recommendation architectures, highlighting their trade-offs in training efficiency, inference efficiency, and ranking performance.
Despite the significant progress made by recent models such as LRURec and HSTU, their ability to effectively capture long-range dependencies still remains somewhat limited.

\begin{table}[t]
\centering
\small
\caption{Summary of major sequential recommendation architectures, comparing their efficiency, ranking performance, and long-sequence modeling capability.}
\label{tab:comparison_main}
\resizebox{\linewidth}{!}{
\begin{tabular}{cccccc}
\toprule
\textbf{Group} & \textbf{Model} & 
\makecell{\textbf{Training} \\ \textbf{Efficiency}} & 
\makecell{\textbf{Inference} \\ \textbf{Efficiency}} & 
\makecell{\textbf{Rec.} \\ \textbf{Performance}} &
\makecell{\textbf{Long-seq.} \\ \textbf{Capability}} \\
\midrule
RNNs & GRU4Rec~\cite{hidasi2015session} & Low & Medium & Medium & Low \\
     & NARM~\cite{li2017neural} & Low & Medium & Medium & Low \\
     & LRURec~\cite{yue2024linear} & High & High & High & Medium \\
\midrule
SARs & SASRec~\cite{kang2018self} & Medium & Low & High & Medium \\
     & BERT4Rec~\cite{sun2019bert4rec} & Medium & Low & High & Medium \\
     & EulerFormer~\cite{tian2024eulerformer} & Medium & Low & High & Medium \\
\midrule
Generative & HSTU~\cite{zhai2024actions} & High & High & High & Medium \\
\bottomrule
\end{tabular}
}
\end{table}

To address these challenges, we propose \textbf{HyenaRec}, a novel sequential recommender. HyenaRec effectively models user interests while maintaining computational efficiency. Its core component is a long convolution whose kernels are parameterized by Legendre orthogonal polynomials, providing a smooth, compact, and numerically stable basis for capturing temporal dependencies. Our motivation stems from two key observations. First, real-world recommendation systems often involve sparse feedback, where conventional operators struggle to maintain stable representations. Second, user interests exhibit multi-scale sequential dependencies, combining slowly evolving global preferences with short-term bursts, requiring modeling mechanisms that can capture both global and local patterns effectively. Legendre polynomial-based long convolutions directly address these challenges, offering stable representations and the capacity to model temporal patterns efficiently. Extensive experiments on multiple real-world datasets demonstrate that HyenaRec consistently outperforms Attention-, Recurrent-, and other baselines in ranking accuracy. It also trains significantly faster, with particularly pronounced advantages in scenarios where multi-scale sequential dependencies are prominent.

The main contributions of this work are summarized as follows:
\begin{itemize}
    \item We propose \textbf{HyenaRec}, the first sequential recommender to adapt sub-quadratic long convolutions into recommendation through Legendre polynomial parameterization.
    \item We show that polynomial kernel parameterization provides a stable and compact representation space, effectively addressing sparse feedback and multi-scale user behaviors.
    \item We empirically validate \textbf{HyenaRec} on multiple real-world benchmark datasets, demonstrating substantial improvements in overall ranking performance and training/inference efficiency, with particularly notable gains in long-sequence recommendation scenarios.

\end{itemize}

\section{Related Work}
\subsection{Sequential Recommendation}
Sequential recommendation aims to capture users’ evolving preferences from their historical interactions and predict the next item.\cite{hidasi2015session, kang2018self, ren2019lifelong, tang2018personalized}
Early studies were largely \textbf{Markov-based approaches}\cite{he2016fusing, rendle2010factorizing}, which modeled item-to-item transitions through first-order or higher-order transition structures. These methods are simple and effective for short-term dependencies, but their memoryless nature makes them inadequate for capturing long-range user interests.
With the rise of deep learning, two major neural architectures emerged. \textbf{CNN-based models}\cite{tang2018personalized, yan2019cosrec} leverage convolutional filters over recent actions to extract strong local patterns and support parallel training, but their fixed receptive fields limit the ability to model long-range dependencies. \textbf{RNN-based models}\cite{hidasi2018recurrent, hidasi2015session, li2017neural, yue2024linear} recurrently encode the entire history into hidden states, enabling efficient online inference. Variants such as LRURec~\cite{yue2024linear} further improve training and inference efficiency on long sequences by leveraging linear recurrence and parallelization techniques. However, issues such as gradient vanishing and the “summary bottleneck” still hinder their ability to retain long-term information in extended sequences.
 However, issues such as gradient vanishing and the “summary bottleneck” hinder their ability to retain long-term information in extended sequences.

To overcome these limitations, \textbf{attention-based}\cite{devlin2019bert, vaswani2017attention} architectures have been widely adopted. Benefiting from the self-attention mechanism, these models can capture both short- and long-range dependencies and have achieved state-of-the-art performance in sequential recommendation\cite{he2021locker, kang2018self, sun2019bert4rec, tian2024eulerformer}. Nevertheless, their quadratic time and memory complexity poses challenges in scenarios involving long interaction histories and large-scale applications. Recent advances such as HSTU~\cite{zhai2024actions} and Mamba4Rec~\cite{liu2024mamba4rec} attempt to alleviate these limitations via efficient state-space formulations.

\subsection{Hyena operator in language Modeling}
Recent advances in sequence modeling have sought to overcome the quadratic complexity of Transformer-style attention\cite{gu2021efficiently, orvieto2023resurrecting, poli2023hyena}.  Within this context, the Hyena operator\cite{poli2023hyena} has emerged as a promising alternative that combines the expressive power of attention with the efficiency of convolutional and state-space formulations.  Unlike traditional attention, which explicitly computes pairwise token interactions, Hyena leverages long convolutional filters and implicit recurrence to capture both short- and long-range dependencies in linear time.  This design allows it to scale to very long sequences, while significantly reducing memory usage.

In language modeling, \textbf{Hyena-based }architectures\cite{nguyen2023hyenadna, massaroli2023laughing, ralambomihanta2024scavenging} demonstrate that high-capacity sequence models do not necessarily require attention mechanisms, providing a new perspective on efficient representation learning. By decoupling sequence length from computational cost, these advanced models show strong potential not only in natural language processing but also in various domains where long-context reasoning is essential, such as recommendation systems and time-series forecasting tasks.

Although the Hyena operator shows promise in language modeling by scaling to long contexts with linear complexity, its direct adoption in recommendation is hindered by two key challenges. The first is the limited representation capacity under sparse user-item interactions. The second is the difficulty of capturing highly diverse, long user behavior sequences. To bridge this gap, we introduce \textbf{HyenaRec}, which augments Hyena with convolutional kernels based on Legendre orthogonal polynomials to model smooth long-term user dependencies. A complementary gating mechanism highlights short-term behavioral bursts. This hybrid design enables HyenaRec to balance global preference evolution with localized user interests, making it better suited for large-scale sequential recommendation under sparse feedback.

\section{Methodology}

\subsection{Problem Formulation}

Let the user’s interaction sequence be $\mathcal{S}=[i_1,\dots,i_T]$ with $i_t\in\mathcal{Y}$, where $\mathcal{Y}$ is the item corpus of size $V=|\mathcal{Y}|$.  
The task is to predict the next item $i_{T+1}$.

A parameter-$\bm\theta$ model $f_{\bm\theta}$ maps $\mathcal{S}$ to logits $\bm s\in\mathbb R^{V}$.
We train by minimising the negative log-likelihood
\begin{equation}
\mathcal L(\bm\theta)=-\log\frac{\exp(s_{i_{T+1}})}{\sum_{j\in\mathcal Y}\exp(s_j)}
\end{equation}
over training pairs $(\mathcal S,i_{T+1})\sim\mathcal D$, using Adam with mini-batches.  
At inference, the top-$k$ items with highest scores are returned and evaluated via Recall@$k$, NDCG@$k$.

\begin{figure*}
    \centering
    \includegraphics[width=\textwidth]{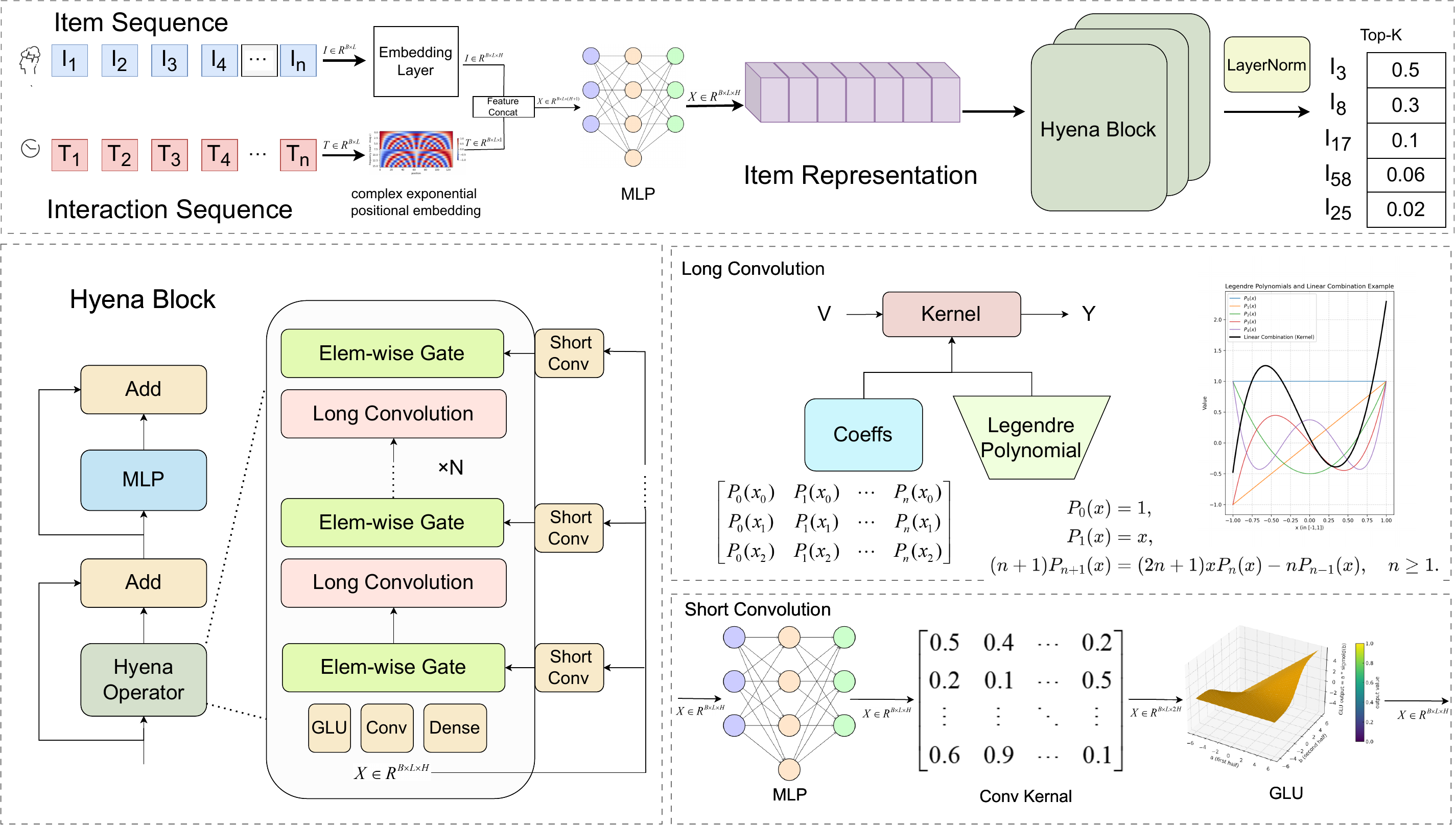}
    \caption{Overall Architecture of \textbf{HyenaRec}, depicting the input \& embedding layer, Hyena-based sequential backbone (where each Hyena Block contains a Hyena Operator that utilizes Legendre Hyena Filters for long-range modeling), and the Top-K prediction layer for next-item recommendation.}
    \label{fig:overall-hyenarec}
\end{figure*}

\subsection{Overall Architecture of HyenaRec}
\label{sec:overall-arch}

The proposed \textbf{HyenaRec} model is designed to efficiently model ultra-long user interaction sequences while maintaining stability under the sparse and noisy conditions commonly encountered in recommendation scenarios. As illustrated in Figure~\ref{fig:overall-hyenarec}, the overall architecture can be divided into three major components: the input \& embedding layer, the Hyena-based backbone for sequential modeling, and the prediction layer for next-item recommendation.

\textbf{Input \& Embedding Layer.}
The model input is a sequence of item identifiers representing user interaction history. 
The embedding layer is composed of two components:

\emph{(1) Item Embedding Lookup.} 
Each item ID is mapped to a $D$-dimensional dense vector through an embedding table, 
optionally followed by a linear projection if the internal dim $\neq D$. 
This produces $\mathbf{E} \in \mathbb{R}^{B \times L \times D}$ for batch size $B$ and sequence length $L$.

\emph{(2) Positional Encoding.} 
When attention-based mixers are used, explicit positional embeddings are added to $\mathbf{E}$, 
as in standard Transformer architectures. 
In the Hyena-based setting, however, positional information is injected implicitly through 
the filter generation process. Specifically, complex exponential embeddings are constructed 
from a normalized temporal grid:
\begin{equation}
t = \frac{1}{L}\,[0,1,\dots,L-1],
\end{equation}
and for frequency $f=1,\dots,F$ the complex exponential basis is
\begin{equation}
z_f(t) = e^{-i \,\frac{2\pi f}{L} \, t}.
\end{equation}
The final positional embedding at time step $t$ is then defined as
\begin{equation}
\mathbf{p}(t) = \bigl[t, \; \Re(z_1(t)),\dots,\Re(z_F(t)), \; \Im(z_1(t)),\dots,\Im(z_F(t)) \bigr],
\end{equation}
which is provided to the Hyena filter generator rather than directly added to the item embeddings. 
This separation ensures that temporal structure is modeled consistently with the chosen backbone.

\textbf{Hyena-based Backbone.}
At the core of HyenaRec lies a stack of $N$ sequential \emph{Blocks}, each consisting of a mixer layer, a feed-forward MLP, residual connections, and normalization. The mixer layer is instantiated as a \textbf{Hyena Operator} (Section~\ref{sec:hyena-op}), augmented with the \textbf{LegendreHyenaFilter} (Section~\ref{hyenafilter}) to ensure stable kernel generation. Specifically, each block maps an input sequence $\mathbf{H}^{(l)} \in \mathbb{R}^{B \times L \times D}$ into an output $\mathbf{H}^{(l+1)}$ of the same shape:
\begin{equation}
\mathbf{H}^{(l+1)} = \mathrm{Block}\bigl(\mathbf{H}^{(l)}\bigr), \qquad l=1,\dots,N.
\end{equation}
The Hyena mixer combines (i) a short path implemented by lightweight convolutions and adaptive gating, and (ii) a long path realized by channel-wise depthwise convolutions with Legendre-polynomial filters. This design inherits Hyena’s efficient long-range modeling power while providing the strong inductive bias necessary for recommendation sequences.

\textbf{Prediction Layer.}
The backbone produces a sequence of hidden states $\mathbf{H}^{(N)} \in \mathbb{R}^{B \times L \times D}$. Following standard practice in sequential recommendation, we directly project the final hidden states onto the item vocabulary space via \emph{weight tying} with the input embedding matrix. Concretely, the logits are computed as
\begin{equation}
\mathbf{z} = \mathbf{H}^{(N)} W^\top + \mathbf{b}, \qquad W \in \mathbb{R}^{|\mathcal{V}| \times D},
\end{equation}
where $W$ shares parameters with the item embedding table and $\mathbf{b}$ is a learnable bias term. The resulting logits are normalized with a softmax function to produce the next-item probability distribution. This design avoids additional decoder parameters while ensuring efficient and consistent representation learning.

\textbf{Overall Workflow.}
Summarizing the above, HyenaRec processes a user’s historical sequence as follows: 
(1) mapping item identifiers to dense embeddings;
(2) embeddings are passed through multiple Hyena-based blocks that hierarchically model both local patterns and long-range dependencies; 
(3) the resulting hidden states are projected through the prediction layer into a probability distribution over candidate items. 
This architecture combines the near-linear time complexity of Hyena operators with the stability of Legendre polynomial parameterization, provides an effective architecture for ultra-long sequential recommendation.

\subsection{Hyena Operator for Sequential Recommendation}
\label{sec:hyena-op}

Modeling long-range dependencies is central to sequential recommendation, where user histories may span hundreds or thousands of steps. Attention-based decoders achieve strong accuracy but scale quadratically in sequence length, which is impractical for long histories. To mitigate this, we build on the \textbf{Hyena operator} family, originally proposed for efficient long-context modeling in language tasks. Hyena replaces explicit pairwise attention with implicit long convolutions parameterized by structured filter functions, enabling attention-like expressivity at sub-quadratic (often near-linear) cost.

We summarise the Hyena-style computation at a level that is sufficient for understanding our adaptations. Let \(\mathbf{X}\in\mathbb{R}^{B\times L\times D}\) denote a batch of input embeddings (batch size \(B\), length \(L\), embedding dim \(D\)). Hyena first expands channels via a learnable projection:
\begin{equation}
\mathbf{U} = \mathbf{X}W_{\mathrm{in}}, \qquad W_{\mathrm{in}}\in\mathbb{R}^{D\times D_{\mathrm{exp}}},
\end{equation}
where \(D_{\mathrm{exp}}=D\cdot (O+1)\) depends on the recurrence order \(O\). The expanded tensor \(\mathbf{U}\) is routed into two main computational paths: a short, local path via lightweight pointwise/1D convolutions and adaptive gating, and a long path realised by depth-wise convolutions with per-channel filters. 
The short-path output is split into \(O\) modulation streams and an initial long-path state, each stream modulating the corresponding stage of the depth-wise convolution:

\begin{equation}
\bigl\{ \mathbf{X}^{(0)}, \mathbf{X}^{(1)},\dots,\mathbf{X}^{(O-1)}, \mathbf{V}^{(0)} \bigr\}
=
\mathrm{ShortPath}(\mathbf{U}),
\end{equation}
where each \(\mathbf{X}^{(o)}\in\mathbb{R}^{B\times L\times d}\) explicitly serves as a data-dependent modulation signal for the corresponding stage of the long-path computation. 
The term \(\mathbf{V}^{(0)}\) provides the initial hidden state that is convolved throughout the long path.

At each processing stage \(o=1,\dots,O\!-\!1\), the long-path hidden state is dynamically modulated and filtered:

\begin{equation}\label{eq:hyena-step}
\mathbf{V}^{(o)} \;=\; \mathrm{Conv}_{\mathrm{dw}}\!\bigl(\;\mathbf{V}^{(o-1)} \odot \mathbf{X}^{(o)}\;,\; k^{(o)} \;\bigr),
\end{equation}
where each \(\mathbf{X}^{(o)}\in\mathbb{R}^{B\times L\times d}\) modulates the corresponding stage of the long-path convolution (here \(d = D_{\mathrm{ch}}\)), and \(\mathbf{V}^{(0)}\) is the initial state to be convolved. Here, \(\odot\) denotes element-wise (channel-wise, time-aligned) multiplication and \(\mathrm{Conv}_{\mathrm{dw}}\) is a causal depth-wise 1D convolution applied independently per channel using the per-stage filter \(k^{(o)}\in\mathbb{R}^{D_{\mathrm{ch}}\times L}\). The final output is obtained by a last modulation and a projection:
\begin{equation}
\mathbf{Y} \;=\; \bigl(\mathbf{V}^{(O-1)} \odot \mathbf{X}^{(0)}\bigr) W_{\mathrm{out}},
\end{equation}
with \(W_{\mathrm{out}}\in\mathbb{R}^{d\times D}\) restoring the original embedding width. By composing multiple stages of gated long convolutions, the effective temporal receptive field expands progressively with the number of stages \(O\), enabling Hyena to capture dependencies at multiple time scales without explicitly constructing \(L\times L\) attention matrices.

A key design choice in Hyena is how the per-channel, per-stage filters \(k^{(o)}\) are parameterized. 
While the original Hyena literature favors \emph{implicit} parameterizations---using a compact generator 
(e.g., an MLP with Fourier features) to produce the kernel samples---our approach adopts an \emph{explicit} 
polynomial expansion. Specifically, each kernel is represented as a linear combination of Legendre bases:
\begin{equation}
k^{(o)}(t) \;=\; \sum_{n=0}^{K-1} c^{(o)}_{n}\, P_{n}\!\left(\mathrm{linspace}(-1,1,L)[t]\right), 
\qquad t=0,\dots,L-1,
\end{equation}
where \(P_{n}(\cdot)\) denotes the \(n\)-th Legendre polynomial and \(c^{(o)}_{n}\) are learnable coefficients. 
This formulation preserves the parameter efficiency of implicit generators (\(O(DK)\) parameters with 
\(K \ll L\)), while offering \textbf{exact orthogonality}, controlled boundary behavior, and numerical stability 
that mitigate overfitting on sparse interactions.This effectively enforces a low-pass filtering effect, suppressing noisy high-frequency fluctuations in user behavior.

Convolutions with these kernels are still performed in the frequency domain via FFT, incurring 
\(\mathcal{O}(BDL \log L)\) computational complexity. Compared to typical MLP-based kernel generators, the Legendre expansion provides a structured and highly regularized parameterization, ensuring both scalability and kernel-level interpretability in large-scale long-sequence recommendation scenarios.

While effective in dense sequence domains, common implicit filter parameterizations exhibit limitations in recommendation settings. 
User–item interaction sequences are typically sparse and noisy, causing flexible generators $g_\theta$ to overfit spurious high-frequency patterns. 
Moreover, large variations in sequence length can introduce numerical instability and interpolation artifacts when filters are rescaled across different $L$, especially for spectral or Fourier-based designs that assume periodic structure. 
Although Hyena’s hierarchical stages offer multi-resolution modeling in principle, the original implicit generators do not explicitly enforce the smooth, low-frequency inductive biases required for robust long-range modeling under sparse supervision.

In \textbf{HyenaRec}, we retain the efficient architectural scaffold of the Hyena Operator, including channel expansion, short-path gating, depth-wise long convolutions, and hierarchical composition. 
We replace the canonical implicit filter generator with a structured, low-dimensional projection onto an orthogonal polynomial basis.
Concretely, instead of \(k^{(o)} = g_\theta^{(o)}(\cdot)\), each per-channel filter is parameterized as a linear combination of \(K\) global Legendre polynomials (Section~\ref{hyenafilter}). 
This design improves numerical stability, introduces stronger implicit regularization in sparse-data regimes, and enables explicit control over the smoothness of long convolution kernels. 
Overall, this modification preserves Hyena’s computational efficiency while adapting its inductive bias to the statistical characteristics of recommendation data.

\begin{figure*}
    \centering
    \includegraphics[width=\linewidth]{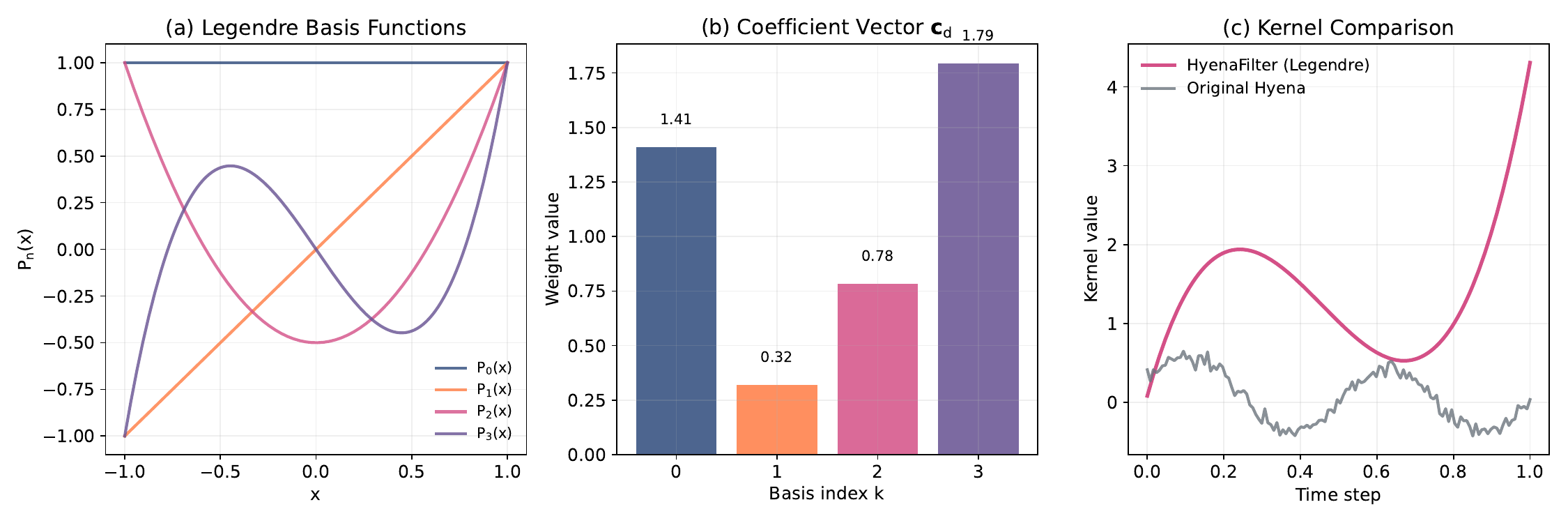}
    \caption{Visualization of the LegendreHyenaFilter. (a) Legendre polynomial basis functions \(P_n(x)\)(\(n = 0,1,2,3\)) exhibit orthogonality and stable boundary behavior. (b) A coefficient vector\(\mathbf{c}_d\) (for a single channel) linearly weights these bases to construct channel-specific convolutional kernels. (c) Comparison between kernels from LegendreHyenaFilter (smoother, more stable) and original Hyena, emphasizing suitability for sparse, variable-length recommendation sequences.
}
    \label{fig:LegendreHyenaFilter}
\end{figure*}

\subsection{LegendreHyenaFilter: Smooth Orthogonal Projection for Long Convolutions}
\label{hyenafilter}

A core component of the Hyena operator is the implicit long-range convolution, where each channel's filter is generated via a compact parametrization rather than stored explicitly. While the original Hyena parameterizes long convolution kernels implicitly using small position-conditioned networks in NLP tasks, 
directly applying this approach to recommendation sequences faces several challenges. 
User–item interactions are extremely sparse, sequence lengths vary widely, and contextual windows can be very long, 
which together can lead to overfitting, numerical instability, and gradient-related issues during training.

To address these limitations while preserving the sub-quadratic efficiency of Hyena, we propose the \textbf{LegendreHyenaFilter}. Figure~\ref{fig:LegendreHyenaFilter} illustrates the overall structure and key components of this filter, including the polynomial bases, channel-specific coefficient weighting, and comparison with original Hyena kernels. This approach constrains each channel's convolutional kernel to lie within a low-dimensional subspace spanned by the first $K$ Legendre polynomials, defined on the interval $x \in [-1,1]$ via Rodrigues' formula:
\begin{equation}
P_n(x) = \frac{1}{2^n n!} \frac{d^n}{dx^n}\left[(x^2-1)^n\right], \quad n=0,\dots,K-1.
\end{equation}

The Legendre polynomial basis offers several advantageous properties for modeling long-range dependencies in recommendation sequences (Figure~\ref{fig:LegendreHyenaFilter}a). First, their orthogonality under the uniform measure ensures stable extrapolation across variable sequence lengths while avoiding the periodic artifacts typical of Fourier bases. Second, the well-controlled boundary behavior---with $|P_n(\pm 1)| = 1$ for all $n$---provides predictable extreme values, mitigating potential signal explosion at sequence boundaries.

To maintain stability in deep networks, we incorporate causal normalization by $\ell_1$-normalizing each channel's kernel after linearly combining the polynomial bases:

\begin{equation}
\sum_{t=0}^{L-1} \bigl| k_t^{(o)} \bigr| = 1, \quad \forall o,
\end{equation}
which prevents exponential growth or decay of signals as they propagate through multiple Hyena layers, thereby stabilizing both forward passes and gradient flow during backpropagation. This constraint acts similarly to residual normalization in RNNs, preventing gradient explosion across stacked Hyena layers.

For efficient computation of Legendre polynomials, we leverage the classic three-term recurrence relation:
\begin{equation}
(n+1)P_{n+1}(x) = (2n+1)x P_n(x) - n P_{n-1}(x),
\end{equation}
enabling highly efficient evaluation of the entire polynomial basis in $\mathcal{O}(KL)$ operations while fully maintaining seamless compatibility with automatic differentiation frameworks.

The complete kernel construction process proceeds as follows. Let $\mathbf{P} \in \mathbb{R}^{K \times L}$ denote the basis matrix obtained by evaluating the Legendre polynomials on a uniform temporal grid, and $\mathbf{C} \in \mathbb{R}^{D \times K}$ represent the learnable coefficient matrix. The channel-wise convolutional kernels are then given by:
\begin{equation}
\mathbf{K} = \frac{\mathbf{C}\mathbf{P}}
{\|\mathbf{C}\mathbf{P}\|_{1,\text{channel}}}
\in \mathbb{R}^{D \times L},
\end{equation}
where each channel is constructed via a linear combination of bases weighted by coefficients (Figure~\ref{fig:LegendreHyenaFilter}b), and the normalization operation ensures the sum-to-one constraint described above.

A critical design consideration is the selection of an appropriate basis size $K$ that balances expressiveness with parameter efficiency. To inform this choice quantitatively, we define the cumulative squared energy captured by the first $K$ basis functions:
\begin{equation}
E_K = \frac{\sum_{n=0}^{K-1} \|\mathbf{c}_n \circ \mathbf{p}_n\|_2^2}{\sum_{n=0}^{N-1} \|\mathbf{c}_n \circ \mathbf{p}_n\|_2^2},
\end{equation}
where $N$ is a sufficiently large reference value. Through empirical analysis on representative recommendation sequences, we set $K=64$ to ensure $E_K \ge 0.95$, thereby preserving the majority of relevant temporal patterns while maintaining computational efficiency.

The final depthwise convolution operation applies the constructed kernels to the input sequence $\mathbf{X} \in \mathbb{R}^{B \times L \times D}$:
\begin{equation}
\mathbf{Y} = \mathbf{X} \ast \mathbf{K},
\end{equation}
yielding smoother, more stable kernels compared with the original Hyena (Figure~\ref{fig:LegendreHyenaFilter}c), and preserving the sub-quadratic complexity of the original operator while endowing it with enhanced stability and regularization properties tailored to recommendation data.

In summary, the LegendreHyenaFilter extends the original Hyena through three key innovations: 
(1) a structured low-dimensional parameterization that reduces parameters while preserving expressiveness; 
(2) a stability-oriented design that combines orthogonal polynomial projection with per-channel kernel normalization to control signal and gradient propagation; and 
(3) an energy-based criterion for basis selection that balances modeling capacity and computational efficiency. 
Collectively, these contributions yield smooth and globally-aware convolutional kernels that are well suited for modeling long-range dependencies in sparse, variable-length recommendation sequences in practice.

\subsection{Discussion}

\textbf{Why Legendre Polynomials?} 
In recommendation, user interaction distributions are highly non-periodic and sparse; thus, Legendre bases naturally align with this irregular temporal structure. Legendre polynomials provide a unique combination of numerical stability, computational efficiency, and inherent regularization properties that make them particularly suitable for sparse sequential recommendation. Unlike Fourier bases, which imposes artificial periodicity on non-periodic user behaviors\cite{2000Chebyshev}, or Chebyshev polynomials, which may amplify numerical instabilities at boundaries\cite{mason2002chebyshev}, Legendre polynomials maintain controlled boundary values ($|P_n(\pm 1)| = 1$) while offering exact orthogonality under the uniform measure. This orthogonal structure ensures minimal redundancy across polynomial orders, providing implicit regularization that prevents overfitting to sparse interaction patterns. Furthermore, the three-term recurrence relation enables $\mathcal{O}(KL)$ basis generation that is both efficient and numerically stable, avoiding catastrophic cancellation issues of high-order polynomial evaluation. The smooth decay of high-order coefficients naturally filters high-frequency noise in user interactions, and the resulting structured filters provide inductive biases for capturing long-term dependencies. As verified in our ablation study (Section~\ref{tab:ablation}), this combination of properties allows HyenaRec to model complex sequential dynamics without sacrificing stability or scalability.

\textbf{Can HyenaRec handle long sequences efficiently?} 
HyenaRec achieves sub-quadratic scaling with respect to sequence length, offering significant advantages over RNN- and attention-based architectures, both theoretically and empirically. The model's parameters decompose into embeddings ($\mathcal{O}(|\mathcal{V}|D)$), backbone projections ($\mathcal{O}(ND^2)$), and lightweight Legendre coefficients ($\mathcal{O}(NDK)$), with $K \ll D$ ensuring parameter efficiency. Each layer requires $\mathcal{O}(BLD(D + \log L))$ operations—dominated by MLP projections and Hyena convolutions—allowing the model to scale efficiently to ultra-long sequences. This compares favorably to attention-based models ($\mathcal{O}(BL^2D)$) and optimized RNNs ($\mathcal{O}(BLD^2)$) when $D > \log L$. Crucially, HyenaRec maintains only $\mathcal{O}(LD)$ memory during training—versus $\mathcal{O}(L^2)$ for attention—permitting long sequences without accuracy loss. For inference, the convolutional formulation supports $\mathcal{O}(D)$ incremental computation per step, enabling real-time serving in production. In addition, the hierarchical structure of the Hyena operator enables interactions across multiple temporal scales to be modeled efficiently, providing robust representation of complex user behavior patterns and improving stability under sparse feedback.
Section~\ref{experiments} demonstrates these advantages, highlighting HyenaRec’s scalability, robustness, and practical suitability for industrial-scale sequential recommendation.

\begin{table}[t]
\centering
\caption{Dataset statistics after preprocessing.}
\resizebox{\linewidth}{!}{% 
\begin{tabular}{cccccc}
\hline
\textbf{Datasets} & \textbf{Users} & \textbf{Items} & \textbf{avg.\,Length} & \textbf{Interact.} & \textbf{Sparsity} \\
\hline
ML-1M  & 6{,}040   & 3{,}416   & 165.5 & 1\,$M$   & 95.2\,\% \\
Steam  & 334{,}730 & 13{,}047  & 11.0  & 3.7\,$M$ & 99.96\,\% \\
Video  & 31{,}013  & 23{,}715  & 9.3   & 287\,$K$ & 99.9\,\% \\
XLong  & 69{,}691  & 2{,}122{,}932 & 958.8 & 66.8\,$M$ & 99.95\,\% \\
\hline
\end{tabular}%
}
\label{tab:comparison}
\end{table}

\begin{table*}
\centering
\caption{Performance comparison of sequential recommendation models across datasets. HyenaRec outperforms baselines.}

\resizebox{\textwidth}{!}{
\begin{tabular}{llcccccccccc}
\toprule

\multirow{2}{*}{\textbf{Dataset}} & \multirow{2}{*}{\textbf{Metric}} & \textbf{GRU4Rec} & \textbf{NARM} & \textbf{SASRec} & \textbf{BERT4Rec} & \textbf{EulerFormer} & \textbf{HSTU} & \textbf{LRURec}  & \multirow{2}{*}{\textbf{HyenaRec}} & \multirow{2}{*}{\textbf{Improv.}} \\
& & [CIKM 2018] & [CIKM 2017] & [ICDM 2018]  & [CIKM 2019] & [SIGIR 2024]  & [ICML 2024] & [CIKM 2024]  &  & \\
\midrule
\multirow{4}{*}{ML-1M}
& NDCG@10 ↑   & 0.15943 & 0.15310 & 0.18293 & 0.16377 & 0.16016 & 0.18607 & \underline{0.19065} & \textbf{0.20007} & 4.94\% \\
& NDCG@20 ↑   & 0.18735 & 0.17774 & 0.21232 & 0.19118 & 0.18937 & 0.21118 &  \underline{0.21767} & \textbf{0.22878} & 5.10\% \\
& Recall@10 \ ↑ & 0.28317 & 0.27290 & 0.31441 & 0.30894 & 0.29257 & 0.31349 & \underline{0.32402} & \textbf{0.33471} & 3.30\% \\
& Recall@20 \ ↑ & 0.39378 & 0.37066 & 0.43084 & 0.41738 & 0.40368 & 0.41298   & \underline{0.43118} & \textbf{0.44858} & 4.04\% \\

\midrule
\multirow{4}{*}{Steam}
& NDCG@10 ↑   & 0.06512 & 0.06480  & 0.06759 & 0.06433 & 0.06825 & \underline{0.07094} & 0.06934 & \textbf{0.07238} & 2.02\% \\
& NDCG@20 ↑   & 0.08050 & 0.08000  & 0.08298 & 0.08117 & 0.07916 & \underline{0.08736} & 0.08508 & \textbf{0.08925} & 2.16\% \\
& Recall@10 \ ↑ & 0.12204 & 0.12166  & 0.12597 & 0.12188 & 0.12037 & \underline{0.13239} & 0.12831 & \textbf{0.13513} & 2.07\% \\
& Recall@20 \ ↑ & 0.18319 & 0.18199  & 0.18716 & 0.18245 & 0.18416 & \underline{0.19764} & 0.19082  & \textbf{0.20222} & 2.32\% \\

\midrule
\multirow{4}{*}{Video}
& NDCG@10 ↑   & 0.05081 & 0.05329  & 0.05829 & 0.05600 & 0.05473 & \underline{0.06320} & 0.06198 &\textbf{0.06436} & 1.84\% \\
& NDCG@20 ↑   & 0.06200 & 0.06512  & 0.07111 & 0.06958 & 0.06834 & \underline{0.07624} & 0.07547 & \textbf{0.07713} & 1.17\% \\
& Recall@10 \ ↑ & 0.09512 & 0.09777  & 0.10910 & 0.11252 & 0.10635 & 0.11286 & \underline{0.11337} & \textbf{0.11620} & 2.50\% \\
& Recall@20 \ ↑ & 0.13959 & 0.14488  & 0.15989 & 0.16640 & 0.15231 & 0.16470 & \underline{0.16705} & \textbf{0.16720} & 0.01\% \\
\midrule
\multirow{4}{*}{XLong}
& NDCG@10 ↑   & 0.20950 & 0.25375 & 0.30949 & 0.23429 & 0.27164 & 0.28728 & \underline{0.34867} &\textbf{0.41021} & 17.65\% \\
& NDCG@20 ↑   & 0.23729 & 0.29583 & 0.33531 & 0.27293 & 0.30561 & 0.31272 & \underline{0.37174} & \textbf{0.42314} & 13.83\% \\
& Recall@10 \ ↑ & 0.35638 & 0.42482 & 0.49348 & 0.41536 & 0.43825 & 0.45722 & \underline{0.52688} & \textbf{0.54178} & 2.83\% \\
& Recall@20 \ ↓ & 0.46633 & 0.51295 & 0.59527 & 0.56860 & 0.54062 & 0.55766 & \textbf{0.61800} & \underline{0.60281} & -2.46\% \\

\bottomrule
\end{tabular}
}

\label{tab:compact_results}

\end{table*}

\section{Experiments}
\label{experiments}
\subsection{Experimental Setup}
\textbf{Datasets}. We evaluate HyenaRec on four widely used and publicly available datasets: ML-1M~\cite{harper2015movielens}, Steam~\cite{kang2018self}, Video~\cite{he2016ups, mcauley2015image}, and Xlong~\cite{ren2019lifelong}. These datasets cover different domains, including movies, games, and videos, and vary in interaction density and sequence length, offering a diverse benchmark for sequential recommendation. Table~\ref{tab:comparison} summarizes their statistics.

\textbf{Data Preprocessing.} Following standard practice, we adopt the leave-one-out evaluation protocol: for each user, the last interaction is used for testing, the second-to-last for validation, and the remaining for training. We further filter items with fewer than five interactions and users with sequences shorter than two to ensure data quality and consistency.

\textbf{Baseline Methods.}
We compare HyenaRec with representative state-of-the-art sequential recommendation models spanning recurrent-based, attention-based, and linear paradigms. Detailed descriptions of all baselines are provided in Appendix~\ref{baselinedetail}. The baselines include \textbf{GRU4Rec}~\cite{hidasi2015session,hidasi2018recurrent}, an RNN-based model for short-term sequential patterns; \textbf{NARM}~\cite{li2017neural}, which augments RNNs with attention to capture users’ main intents; \textbf{SASRec}~\cite{kang2018self}, a Transformer-based model with causal self-attention; \textbf{BERT4Rec}~\cite{sun2019bert4rec}, a bidirectional Transformer trained with a cloze-style objective; \textbf{EulerFormer}~\cite{tian2024eulerformer}, which employs complex-valued attention for joint semantic–positional encoding; \textbf{HSTU}~\cite{zhai2024actions}, a hierarchical sequential transducer for generative recommendation; and \textbf{LRURec}~\cite{yue2024linear}, a linear recurrent model enabling efficient long-context modeling.

In addition to the standard baselines, we include \textbf{Mamba4Rec}~\cite{liu2024mamba4rec}, a recent selective state-space model, to compare structured convolution with selective recurrence under long-sequence settings. We further evaluate an atrous convolution variant of our model (\textbf{AtrousHyena}) to study the effect of multi-scale receptive fields. Results are reported in Appendix~\ref{mamba4reccompare} and Appendix~\ref{atroushyenamodel}.

\textbf{Evaluation Metrics.} We employ four widely-adopted ranking metrics for performance evaluation: NDCG@10, Recall@10, NDCG@20, and Recall@20. NDCG emphasizes the quality of top-ranked items by considering positional relevance, while Recall measures the model's capability to capture relevant items within the recommended list. These metrics provide complementary perspectives on recommendation quality and coverage.

\textbf{Implementation Details.} We use official implementations from the original papers. Models are trained using cross-entropy loss with the AdamW optimizer at a fixed learning rate of 1e-3. The batch size is 128 (32 for XLong), with a maximum of 500 epochs. Validation is performed every 500–2000 steps, following LRU\cite{yue2024linear}. We employ early stopping if Recall@10 does not improve for 10 consecutive evaluations. Hyperparameter tuning grids search weight decay {0, 1e-8, 1e-6, 1e-4, 1e-2} and dropout {0, 0.2, 0.4, 0.6, 0.8}; other settings follow the original defaults. See Appendix~\ref{implementation} for full details.

\begin{table*}[t]
\centering
\caption{Ablation results of HyenaRec and its variants on four datasets. Each entry reports NDCG and Recall at cutoff 10 and 20.}
\resizebox{0.95\textwidth}{!}{
\renewcommand{\arraystretch}{1.0} % 行高
\setlength{\tabcolsep}{4pt}       % 列间距
\begin{tabular}{llcccc}
\toprule
\multirow{2}{*}{\textbf{Variants}} & \multirow{2}{*}{\textbf{Metric}} &
\textbf{ML-1M} & \textbf{Steam} & \textbf{Video} & \textbf{XLong} \\
 & & \textbf{NDCG $\uparrow$ / Recall $\uparrow$} &
 \textbf{NDCG $\uparrow$ / Recall $\uparrow$} &
 \textbf{NDCG $\uparrow$ / Recall $\uparrow$} &
 \textbf{NDCG $\uparrow$ / Recall $\uparrow$} \\
\midrule

\multirow{2}{*}{HyenaRec}
& @10  &\textbf{0.20007} / \textbf{0.33471} & \textbf{0.07238} / \textbf{0.13513} & \textbf{0.06436} / \textbf{0.11620} & \textbf{0.41021 }/ \textbf{0.54178} \\
& @20  &\textbf{0.22878} / \textbf{0.43118} & \textbf{0.08925} / \textbf{0.20222} & 0.07713/ \textbf{0.16720} & \textbf{0.42314}/  \textbf{0.60281 }\\
\midrule
\multirow{2}{*}{(1) HyenaRec w/o PK}
 & @10 & 0.19174 / 0.3251 & 0.07112 / 0.13280 & 0.06177 / 0.11082 & 0.40552 / 0.48575 \\
 & @20 & 0.21760 / 0.42759 & 0.08738 / 0.19748 & 0.07422 / 0.16038 & 0.41209 / 0.5117 \\
\midrule
\multirow{2}{*}{(2) HyenaRec w/o GLU}
 & @10 & 0.19026 / 0.32229 & 0.06855 / 0.12936 & 0.06338 / 0.11329 & 0.40865 / 0.52341 \\
 & @20 & 0.21730 / 0.42943 & 0.08499 / 0.19474 & \textbf{0.07768 }/ 0.16677 & 0.41936 / 0.58263 \\
\midrule
\multirow{2}{*}{(3) HyenaRec (Fourier)}
 & @10 & 0.19270 / 0.32713 & 0.06955 / 0.13041 & 0.06306 / 0.11375 & 0.40121 / 0.51899 \\
 & @20 & 0.21782 / 0.42689 & 0.08565 / 0.19450 & 0.07623 / 0.16613 &  0.41966/ 0.57333\\
\midrule
\multirow{2}{*}{(4) HyenaRec (Chebyshev)}
 & @10 & 0.19545 / 0.32651 & 0.06902 / 0.13036 & 0.06275 / 0.11271 & 0.40199 / 0.51333 \\
 & @20 & 0.22110 / 0.4284 & 0.08537 / 0.19541 & 0.07584 / 0.16478 & 0.41789 / 0.57625 \\
\bottomrule
\end{tabular}
}

\label{tab:ablation}
\end{table*}

\begin{figure}
    \centering
    \includegraphics[width=\linewidth]{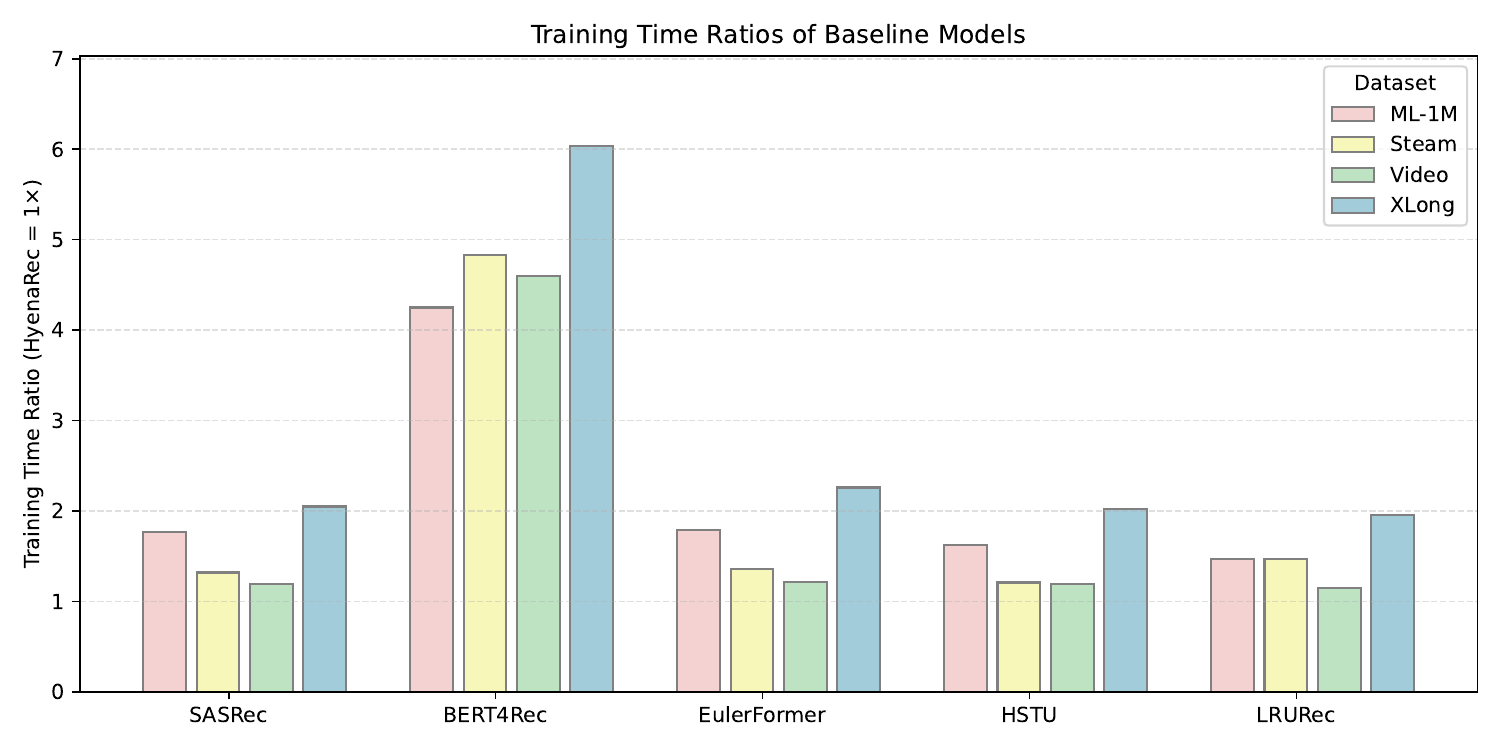}
    \caption{Training time ratios (TTR) of baseline models relative to HyenaRec, which is normalized to 1×. Each bar shows the relative training cost of a model on a given dataset.}
    \label{fig:placeholder}
\end{figure}

\subsection{Overall Performance}
\subsubsection{Model Performance}
Table \ref{tab:compact_results} compares our HyenaRec model with representative sequential recommenders, including GRU4Rec\cite{hidasi2018recurrent}, NARM\cite{li2017neural}, SASRec\cite{kang2018self}, BERT4Rec\cite{sun2019bert4rec}, EulerFormer\cite{tian2024eulerformer}, HSTU\cite{zhai2024actions}, and LRURec\cite{yue2024linear}. Bold numbers indicate the best performance, while underlined numbers represent the second-best results. Across all datasets (ML-1M, Steam, Video, and XLong), HyenaRec consistently achieves the top or second-best NDCG and Recall.

On short-to-medium sequences (ML-1M and Steam), HyenaRec outperforms the strongest baseline (LRURec) by 3–5\% on average in both NDCG@10 and Recall@20, showing its effectiveness in capturing fine-grained sequential dependencies. On long sequences, especially XLong, the gains are more pronounced, reaching 17.65\% improvement in NDCG@10 and 13.83\% in NDCG@20, demonstrating superior handling of ultra-long interaction histories where self-attention–based methods often struggle to scale efficiently. Overall, these results highlight HyenaRec’s ability to balance expressiveness and scalability, outperforming existing state-of-the-art models across diverse recommendation scenarios.

\subsubsection{Time Efficiency}
We evaluate computational efficiency using the \textit{training time ratio} (TTR) relative to HyenaRec, normalized to 1×. As shown in Figure~\ref{fig:placeholder}, all baselines require more training time across datasets (ML-1M, Steam, Video, XLong). For example, on ML-1M, SASRec and BERT4Rec take 1.77× and 4.25× longer, respectively, while HSTU requires 1.62×, highlighting HyenaRec’s superior efficiency. This gain stems from its convolution-based hierarchical representation and sub-quadratic sequence mixing.

\subsubsection{Ablation Study}
Table \ref{tab:ablation} presents the ablation results of HyenaRec and its variants. The full model consistently achieves the best or near-best performance in terms of NDCG and Recall at both cutoffs (10 and 20), validating our architectural design.  

Removing the Polynomial Kernel (PK) module leads to noticeable performance drops (e.g., Recall@20 on Steam: 0.20222 → 0.19748), indicating PK’s critical role in capturing high-order interactions. Similarly, ablating the GLU gating mechanism results in lower performance on ML-1M and Video, suggesting that GLU effectively filters information to model complex user–item relationships. Replacing the learnable convolution kernel with fixed Fourier or Chebyshev bases also reduces performance, particularly on XLong (Recall@20: 0.60281 → 0.57333).  

Overall, each component contributes meaningfully to HyenaRec’s performance, with the learnable convolution and polynomial kernels being particularly crucial for modeling long-range dependencies in sequential recommendation tasks.

\begin{figure}
    \centering
    \includegraphics[width=0.9\linewidth]{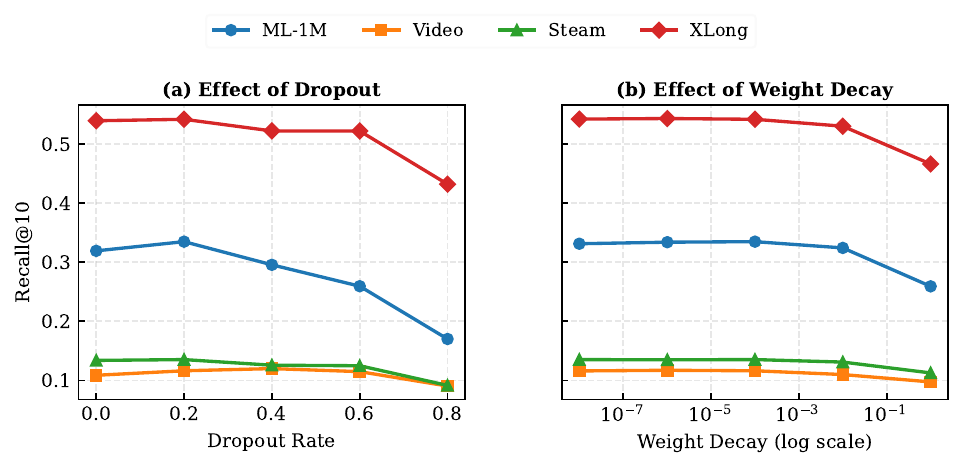}
    \caption{Hyperparameter sensitivity of HyenaRec.}
    \label{fig:sensitivity}
\end{figure}

\subsection{Hyperparameter Sensitivity}
We study HyenaRec's sensitivity to two key training hyperparameters: (i) dropout rate and (ii) weight decay.
Figure~\ref{fig:sensitivity} summarizes results on four datasets (ML-1M, Video, Steam, XLong). We also examine sensitivity to the Legendre filter order $K$ (Appendix~\ref{fig:legendre_sensitivity}), showing stable performance across values.

\textbf{Dropout Rate.} Figure~\ref{fig:sensitivity}(a) shows HyenaRec is stable for dropout 0–0.4, with minor performance fluctuations. Dropout above 0.5 consistently degrades accuracy, indicating excessive stochasticity harms long-term preference modeling. We use 0.2 as default, balancing generalization and expressiveness.

\textbf{Weight Decay.} Figure~\ref{fig:sensitivity}(b) shows HyenaRec prefers mild ${L}^{2}$ regularization, peaking at $\sim 10^{-4}$ and dropping sharply beyond $10^{-2}$, especially on sparse datasets (Steam, XLong). Under-regularization ($\leq 10^{-6}$) causes overfitting on dense datasets (ML-1M, Video). We adopt $10^{-4}$ as default, ensuring stable performance and fast tuning.

\section{Conclusion}

We propose HyenaRec, a sequential recommendation model designed to address efficient modeling of long user behavior sequences. Extensive experiments on multiple datasets demonstrate that HyenaRec consistently outperforms state-of-the-art baselines in ranking performance while maintaining high training and inference efficiency. Our results suggest that structured convolutional operators with orthogonal polynomial parameterization 
offer a promising alternative to attention for long-context sequential modeling.

\begin{acks}
    This work is partially supported by National Natural Science Foundation, China (No.62276196).
\end{acks}

\bibliographystyle{ACM-Reference-Format}
\balance
\bibliography{sample-base}


\begin{thebibliography}{31}

%%% the Web address.

\ifx \showCODEN    \undefined \def \showCODEN     #1{\unskip}     \fi
\ifx \showISBNx    \undefined \def \showISBNx     #1{\unskip}     \fi
\ifx \showISBNxiii \undefined \def \showISBNxiii  #1{\unskip}     \fi
\ifx \showISSN     \undefined \def \showISSN      #1{\unskip}     \fi
\ifx \showLCCN     \undefined \def \showLCCN      #1{\unskip}     \fi
\ifx \shownote     \undefined \def \shownote      #1{#1}          \fi
\ifx \showarticletitle \undefined \def \showarticletitle #1{#1}   \fi
\ifx \showURL      \undefined \def \showURL       {\relax}        \fi
% invisible to TeX
\providecommand\bibfield[2]{#2}
\providecommand\bibinfo[2]{#2}
\providecommand\natexlab[1]{#1}
\providecommand\showeprint[2][]{arXiv:#2}

\bibitem[Boyd et~al\mbox{.}(2000)]%
        {2000Chebyshev}
\bibfield{author}{\bibinfo{person}{John~P. Boyd}, \bibinfo{person}{To Marilyn}, {and} \bibinfo{person}{Paraphrasing T.~S. Eliot}.} \bibinfo{year}{2000}\natexlab{}.
\newblock \showarticletitle{Chebyshev and Fourier Spectral Methods}.
\newblock \bibinfo{journal}{\emph{Dover Publications,}} (\bibinfo{year}{2000}).
\newblock


\bibitem[Devlin et~al\mbox{.}(2019)]%
        {devlin2019bert}
\bibfield{author}{\bibinfo{person}{Jacob Devlin}, \bibinfo{person}{Ming{-}Wei Chang}, \bibinfo{person}{Kenton Lee}, {and} \bibinfo{person}{Kristina Toutanova}.} \bibinfo{year}{2019}\natexlab{}.
\newblock \showarticletitle{{BERT:} Pre-training of Deep Bidirectional Transformers for Language Understanding}. In \bibinfo{booktitle}{\emph{Proceedings of the 2019 Conference of the North American Chapter of the Association for Computational Linguistics: Human Language Technologies, {NAACL-HLT} 2019, Minneapolis, MN, USA, June 2-7, 2019}}. \bibinfo{publisher}{Association for Computational Linguistics}, \bibinfo{pages}{4171--4186}.
\newblock
\href{https://doi.org/10.18653/V1/N19-1423}{doi:\nolinkurl{10.18653/V1/N19-1423}}


\bibitem[Gu and Dao(2023)]%
        {gu2024mamba}
\bibfield{author}{\bibinfo{person}{Albert Gu} {and} \bibinfo{person}{Tri Dao}.} \bibinfo{year}{2023}\natexlab{}.
\newblock \showarticletitle{Mamba: Linear-Time Sequence Modeling with Selective State Spaces}.
\newblock \bibinfo{journal}{\emph{CoRR}}  \bibinfo{volume}{abs/2312.00752} (\bibinfo{year}{2023}).
\newblock
\showeprint[arXiv]{2312.00752}
\href{https://doi.org/10.48550/ARXIV.2312.00752}{doi:\nolinkurl{10.48550/ARXIV.2312.00752}}


\bibitem[Gu et~al\mbox{.}(2022)]%
        {gu2021efficiently}
\bibfield{author}{\bibinfo{person}{Albert Gu}, \bibinfo{person}{Karan Goel}, {and} \bibinfo{person}{Christopher R{\'{e}}}.} \bibinfo{year}{2022}\natexlab{}.
\newblock \showarticletitle{Efficiently Modeling Long Sequences with Structured State Spaces}. In \bibinfo{booktitle}{\emph{Proceedings of the 10th International Conference on Learning Representations, {ICLR} 2022, April 25-29, 2022}}. \bibinfo{publisher}{OpenReview.net}.
\newblock
\urldef\tempurl%
\url{https://openreview.net/forum?id=uYLFoz1vlAC}
\showURL{%
\tempurl}


\bibitem[Harper and Konstan(2016)]%
        {harper2015movielens}
\bibfield{author}{\bibinfo{person}{F.~Maxwell Harper} {and} \bibinfo{person}{Joseph~A. Konstan}.} \bibinfo{year}{2016}\natexlab{}.
\newblock \showarticletitle{The MovieLens Datasets: History and Context}.
\newblock \bibinfo{journal}{\emph{{ACM} Trans. Interact. Intell. Syst.}} \bibinfo{volume}{5}, \bibinfo{number}{4} (\bibinfo{year}{2016}), \bibinfo{pages}{19:1--19:19}.
\newblock
\href{https://doi.org/10.1145/2827872}{doi:\nolinkurl{10.1145/2827872}}


\bibitem[He and McAuley(2016a)]%
        {he2016fusing}
\bibfield{author}{\bibinfo{person}{Ruining He} {and} \bibinfo{person}{Julian~J. McAuley}.} \bibinfo{year}{2016}\natexlab{a}.
\newblock \showarticletitle{Fusing Similarity Models with Markov Chains for Sparse Sequential Recommendation}. In \bibinfo{booktitle}{\emph{Proceedings of the 16th {IEEE} International Conference on Data Mining ({ICDM} 2016), Barcelona, Spain, December 12-15, 2016}}. \bibinfo{publisher}{{IEEE} Computer Society}, \bibinfo{pages}{191--200}.
\newblock
\href{https://doi.org/10.1109/ICDM.2016.0030}{doi:\nolinkurl{10.1109/ICDM.2016.0030}}


\bibitem[He and McAuley(2016b)]%
        {he2016ups}
\bibfield{author}{\bibinfo{person}{Ruining He} {and} \bibinfo{person}{Julian~J. McAuley}.} \bibinfo{year}{2016}\natexlab{b}.
\newblock \showarticletitle{Ups and Downs: Modeling the Visual Evolution of Fashion Trends with One-Class Collaborative Filtering}. In \bibinfo{booktitle}{\emph{Proceedings of the 25th International Conference on World Wide Web, {WWW} 2016, Montreal, Canada, April 11 - 15, 2016}}. \bibinfo{publisher}{{ACM}}, \bibinfo{pages}{507--517}.
\newblock
\href{https://doi.org/10.1145/2872427.2883037}{doi:\nolinkurl{10.1145/2872427.2883037}}


\bibitem[He et~al\mbox{.}(2021)]%
        {he2021locker}
\bibfield{author}{\bibinfo{person}{Zhankui He}, \bibinfo{person}{Handong Zhao}, \bibinfo{person}{Zhe Lin}, \bibinfo{person}{Zhaowen Wang}, \bibinfo{person}{Ajinkya Kale}, {and} \bibinfo{person}{Julian~J. McAuley}.} \bibinfo{year}{2021}\natexlab{}.
\newblock \showarticletitle{Locker: Locally Constrained Self-Attentive Sequential Recommendation}. In \bibinfo{booktitle}{\emph{Proceedings of the 30th {ACM} International Conference on Information and Knowledge Management, {CIKM} 2021, Queensland, Australia, November 1-5, 2021}}. \bibinfo{publisher}{{ACM}}, \bibinfo{pages}{3088--3092}.
\newblock
\href{https://doi.org/10.1145/3459637.3482136}{doi:\nolinkurl{10.1145/3459637.3482136}}


\bibitem[Hidasi and Karatzoglou(2018)]%
        {hidasi2018recurrent}
\bibfield{author}{\bibinfo{person}{Bal{\'{a}}zs Hidasi} {and} \bibinfo{person}{Alexandros Karatzoglou}.} \bibinfo{year}{2018}\natexlab{}.
\newblock \showarticletitle{Recurrent Neural Networks with Top-k Gains for Session-based Recommendations}. In \bibinfo{booktitle}{\emph{Proceedings of the 27th {ACM} International Conference on Information and Knowledge Management, {CIKM} 2018, Torino, Italy, October 22-26, 2018}}. \bibinfo{publisher}{{ACM}}, \bibinfo{pages}{843--852}.
\newblock
\href{https://doi.org/10.1145/3269206.3271761}{doi:\nolinkurl{10.1145/3269206.3271761}}


\bibitem[Hidasi et~al\mbox{.}(2016)]%
        {hidasi2015session}
\bibfield{author}{\bibinfo{person}{Bal{\'{a}}zs Hidasi}, \bibinfo{person}{Alexandros Karatzoglou}, \bibinfo{person}{Linas Baltrunas}, {and} \bibinfo{person}{Domonkos Tikk}.} \bibinfo{year}{2016}\natexlab{}.
\newblock \showarticletitle{Session-based Recommendations with Recurrent Neural Networks}. In \bibinfo{booktitle}{\emph{Proceedings of the 4th International Conference on Learning Representations ({ICLR} 2016), San Juan, Puerto Rico, May 2-4, 2016}}.
\newblock
\urldef\tempurl%
\url{http://arxiv.org/abs/1511.06939}
\showURL{%
\tempurl}


\bibitem[Hu et~al\mbox{.}(2021a)]%
        {DBLP:journals/toit/HuLLS21}
\bibfield{author}{\bibinfo{person}{Kaixi Hu}, \bibinfo{person}{Lin Li}, \bibinfo{person}{Jianquan Liu}, {and} \bibinfo{person}{Daniel Sun}.} \bibinfo{year}{2021}\natexlab{a}.
\newblock \showarticletitle{DuroNet: {A} Dual-robust Enhanced Spatial-temporal Learning Network for Urban Crime Prediction}.
\newblock \bibinfo{journal}{\emph{{ACM} Trans. Internet Techn.}} \bibinfo{volume}{21}, \bibinfo{number}{1} (\bibinfo{year}{2021}), \bibinfo{pages}{24:1--24:24}.
\newblock
\href{https://doi.org/10.1145/3432249}{doi:\nolinkurl{10.1145/3432249}}


\bibitem[Hu et~al\mbox{.}(2021b)]%
        {DBLP:conf/cikm/HuL0LT21}
\bibfield{author}{\bibinfo{person}{Kaixi Hu}, \bibinfo{person}{Lin Li}, \bibinfo{person}{Qing Xie}, \bibinfo{person}{Jianquan Liu}, {and} \bibinfo{person}{Xiaohui Tao}.} \bibinfo{year}{2021}\natexlab{b}.
\newblock \showarticletitle{What is Next when Sequential Prediction Meets Implicitly Hard Interaction?}. In \bibinfo{booktitle}{\emph{Proceedings of the 30th {ACM} International Conference on Information and Knowledge Management, {CIKM} 2021, Queensland, Australia, November 1-5, 2021}}. \bibinfo{publisher}{{ACM}}, \bibinfo{pages}{710--719}.
\newblock
\href{https://doi.org/10.1145/3459637.3482492}{doi:\nolinkurl{10.1145/3459637.3482492}}


\bibitem[Kang and McAuley(2018)]%
        {kang2018self}
\bibfield{author}{\bibinfo{person}{Wang{-}Cheng Kang} {and} \bibinfo{person}{Julian~J. McAuley}.} \bibinfo{year}{2018}\natexlab{}.
\newblock \showarticletitle{Self-Attentive Sequential Recommendation}. In \bibinfo{booktitle}{\emph{Proceedings of the {IEEE} International Conference on Data Mining, {ICDM} 2018, Singapore, November 17-20, 2018}}. \bibinfo{publisher}{{IEEE} Computer Society}, \bibinfo{pages}{197--206}.
\newblock
\href{https://doi.org/10.1109/ICDM.2018.00035}{doi:\nolinkurl{10.1109/ICDM.2018.00035}}


\bibitem[Li et~al\mbox{.}(2017)]%
        {li2017neural}
\bibfield{author}{\bibinfo{person}{Jing Li}, \bibinfo{person}{Pengjie Ren}, \bibinfo{person}{Zhumin Chen}, \bibinfo{person}{Zhaochun Ren}, \bibinfo{person}{Tao Lian}, {and} \bibinfo{person}{Jun Ma}.} \bibinfo{year}{2017}\natexlab{}.
\newblock \showarticletitle{Neural Attentive Session-based Recommendation}. In \bibinfo{booktitle}{\emph{Proceedings of the 2017 {ACM} on Conference on Information and Knowledge Management, {CIKM} 2017, Singapore, November 06 - 10, 2017}}. \bibinfo{publisher}{{ACM}}, \bibinfo{pages}{1419--1428}.
\newblock
\href{https://doi.org/10.1145/3132847.3132926}{doi:\nolinkurl{10.1145/3132847.3132926}}


\bibitem[Liu et~al\mbox{.}(2024)]%
        {liu2024mamba4rec}
\bibfield{author}{\bibinfo{person}{Chengkai Liu}, \bibinfo{person}{Jianghao Lin}, \bibinfo{person}{Jianling Wang}, \bibinfo{person}{Hanzhou Liu}, {and} \bibinfo{person}{James Caverlee}.} \bibinfo{year}{2024}\natexlab{}.
\newblock \showarticletitle{Mamba4Rec: Towards Efficient Sequential Recommendation with Selective State Space Models}.
\newblock \bibinfo{journal}{\emph{CoRR}}  \bibinfo{volume}{abs/2403.03900} (\bibinfo{year}{2024}).
\newblock
\showeprint[arXiv]{2403.03900}
\href{https://doi.org/10.48550/ARXIV.2403.03900}{doi:\nolinkurl{10.48550/ARXIV.2403.03900}}


\bibitem[Mason and Handscomb(2002)]%
        {mason2002chebyshev}
\bibfield{author}{\bibinfo{person}{John~C Mason} {and} \bibinfo{person}{David~C Handscomb}.} \bibinfo{year}{2002}\natexlab{}.
\newblock \bibinfo{booktitle}{\emph{Chebyshev polynomials}}.
\newblock \bibinfo{publisher}{Chapman and Hall/CRC}.
\newblock


\bibitem[Massaroli et~al\mbox{.}(2023)]%
        {massaroli2023laughing}
\bibfield{author}{\bibinfo{person}{Stefano Massaroli}, \bibinfo{person}{Michael Poli}, \bibinfo{person}{Daniel~Y. Fu}, \bibinfo{person}{Hermann Kumbong}, \bibinfo{person}{Rom~N. Parnichkun}, \bibinfo{person}{David~W. Romero}, \bibinfo{person}{Aman Timalsina}, \bibinfo{person}{Quinn McIntyre}, \bibinfo{person}{Beidi Chen}, \bibinfo{person}{Atri Rudra}, \bibinfo{person}{Ce Zhang}, \bibinfo{person}{Christopher R{\'{e}}}, \bibinfo{person}{Stefano Ermon}, {and} \bibinfo{person}{Yoshua Bengio}.} \bibinfo{year}{2023}\natexlab{}.
\newblock \showarticletitle{Laughing Hyena Distillery: Extracting Compact Recurrences From Convolutions}. In \bibinfo{booktitle}{\emph{Proceedings of the 36th Annual Conference on Neural Information Processing Systems, {NeurIPS} 2023, New Orleans, LA, December 10-16, 2023}}.
\newblock
\urldef\tempurl%
\url{http://papers.nips.cc/paper\_files/paper/2023/hash/371355cd42caaf83412c3fbef4688979-Abstract-Conference.html}
\showURL{%
\tempurl}


\bibitem[McAuley et~al\mbox{.}(2015)]%
        {mcauley2015image}
\bibfield{author}{\bibinfo{person}{Julian~J. McAuley}, \bibinfo{person}{Christopher Targett}, \bibinfo{person}{Qinfeng Shi}, {and} \bibinfo{person}{Anton van~den Hengel}.} \bibinfo{year}{2015}\natexlab{}.
\newblock \showarticletitle{Image-Based Recommendations on Styles and Substitutes}. In \bibinfo{booktitle}{\emph{Proceedings of the 38th International {ACM} {SIGIR} Conference on Research and Development in Information Retrieval, Santiago, Chile, August 9-13, 2015}}. \bibinfo{publisher}{{ACM}}, \bibinfo{pages}{43--52}.
\newblock
\href{https://doi.org/10.1145/2766462.2767755}{doi:\nolinkurl{10.1145/2766462.2767755}}


\bibitem[Nguyen et~al\mbox{.}(2023)]%
        {nguyen2023hyenadna}
\bibfield{author}{\bibinfo{person}{Eric Nguyen}, \bibinfo{person}{Michael Poli}, \bibinfo{person}{Marjan Faizi}, \bibinfo{person}{Armin~W. Thomas}, \bibinfo{person}{Michael Wornow}, \bibinfo{person}{Callum Birch{-}Sykes}, \bibinfo{person}{Stefano Massaroli}, \bibinfo{person}{Aman Patel}, \bibinfo{person}{Clayton~M. Rabideau}, \bibinfo{person}{Yoshua Bengio}, \bibinfo{person}{Stefano Ermon}, \bibinfo{person}{Christopher R{\'{e}}}, {and} \bibinfo{person}{Stephen Baccus}.} \bibinfo{year}{2023}\natexlab{}.
\newblock \showarticletitle{HyenaDNA: Long-Range Genomic Sequence Modeling at Single Nucleotide Resolution}. In \bibinfo{booktitle}{\emph{Proceedings of the 36th Annual Conference on Neural Information Processing Systems, {NeurIPS} 2023, New Orleans, LA, December 10-16, 2023}}.
\newblock
\urldef\tempurl%
\url{http://papers.nips.cc/paper\_files/paper/2023/hash/86ab6927ee4ae9bde4247793c46797c7-Abstract-Conference.html}
\showURL{%
\tempurl}


\bibitem[Orvieto et~al\mbox{.}(2023)]%
        {orvieto2023resurrecting}
\bibfield{author}{\bibinfo{person}{Antonio Orvieto}, \bibinfo{person}{Samuel~L. Smith}, \bibinfo{person}{Albert Gu}, \bibinfo{person}{Anushan Fernando}, \bibinfo{person}{{\c{C}}aglar G{\"{u}}l{\c{c}}ehre}, \bibinfo{person}{Razvan Pascanu}, {and} \bibinfo{person}{Soham De}.} \bibinfo{year}{2023}\natexlab{}.
\newblock \showarticletitle{Resurrecting Recurrent Neural Networks for Long Sequences}. In \bibinfo{booktitle}{\emph{Proceedings of the International Conference on Machine Learning, {ICML} 2023, Honolulu, Hawaii, July 23-29, 2023}} \emph{(\bibinfo{series}{Proceedings of Machine Learning Research}, Vol.~\bibinfo{volume}{202})}. \bibinfo{publisher}{{PMLR}}, \bibinfo{pages}{26670--26698}.
\newblock
\urldef\tempurl%
\url{https://proceedings.mlr.press/v202/orvieto23a.html}
\showURL{%
\tempurl}


\bibitem[Poli et~al\mbox{.}(2023)]%
        {poli2023hyena}
\bibfield{author}{\bibinfo{person}{Michael Poli}, \bibinfo{person}{Stefano Massaroli}, \bibinfo{person}{Eric Nguyen}, \bibinfo{person}{Daniel~Y. Fu}, \bibinfo{person}{Tri Dao}, \bibinfo{person}{Stephen Baccus}, \bibinfo{person}{Yoshua Bengio}, \bibinfo{person}{Stefano Ermon}, {and} \bibinfo{person}{Christopher R{\'{e}}}.} \bibinfo{year}{2023}\natexlab{}.
\newblock \showarticletitle{Hyena Hierarchy: Towards Larger Convolutional Language Models}. In \bibinfo{booktitle}{\emph{Proceedings of the International Conference on Machine Learning, {ICML} 2023, Honolulu, Hawaii, July 23-29, 2023}} \emph{(\bibinfo{series}{Proceedings of Machine Learning Research}, Vol.~\bibinfo{volume}{202})}. \bibinfo{publisher}{{PMLR}}, \bibinfo{pages}{28043--28078}.
\newblock
\urldef\tempurl%
\url{https://proceedings.mlr.press/v202/poli23a.html}
\showURL{%
\tempurl}


\bibitem[Ralambomihanta et~al\mbox{.}(2024)]%
        {ralambomihanta2024scavenging}
\bibfield{author}{\bibinfo{person}{Tokiniaina~Raharison Ralambomihanta}, \bibinfo{person}{Shahrad Mohammadzadeh}, \bibinfo{person}{Mohammad Sami~Nur Islam}, \bibinfo{person}{Wassim Jabbour}, {and} \bibinfo{person}{Laurence Liang}.} \bibinfo{year}{2024}\natexlab{}.
\newblock \showarticletitle{Scavenging Hyena: Distilling Transformers into Long Convolution Models}.
\newblock \bibinfo{journal}{\emph{CoRR}}  \bibinfo{volume}{abs/2401.17574} (\bibinfo{year}{2024}).
\newblock
\showeprint[arXiv]{2401.17574}
\href{https://doi.org/10.48550/ARXIV.2401.17574}{doi:\nolinkurl{10.48550/ARXIV.2401.17574}}


\bibitem[Ren et~al\mbox{.}(2019)]%
        {ren2019lifelong}
\bibfield{author}{\bibinfo{person}{Kan Ren}, \bibinfo{person}{Jiarui Qin}, \bibinfo{person}{Yuchen Fang}, \bibinfo{person}{Weinan Zhang}, \bibinfo{person}{Lei Zheng}, \bibinfo{person}{Weijie Bian}, \bibinfo{person}{Guorui Zhou}, \bibinfo{person}{Jian Xu}, \bibinfo{person}{Yong Yu}, \bibinfo{person}{Xiaoqiang Zhu}, {and} \bibinfo{person}{Kun Gai}.} \bibinfo{year}{2019}\natexlab{}.
\newblock \showarticletitle{Lifelong Sequential Modeling with Personalized Memorization for User Response Prediction}. In \bibinfo{booktitle}{\emph{Proceedings of the 42nd International {ACM} {SIGIR} Conference on Research and Development in Information Retrieval, {SIGIR} 2019, Paris, France, July 21-25, 2019}}. \bibinfo{publisher}{{ACM}}, \bibinfo{pages}{565--574}.
\newblock
\href{https://doi.org/10.1145/3331184.3331230}{doi:\nolinkurl{10.1145/3331184.3331230}}


\bibitem[Rendle et~al\mbox{.}(2010)]%
        {rendle2010factorizing}
\bibfield{author}{\bibinfo{person}{Steffen Rendle}, \bibinfo{person}{Christoph Freudenthaler}, {and} \bibinfo{person}{Lars Schmidt{-}Thieme}.} \bibinfo{year}{2010}\natexlab{}.
\newblock \showarticletitle{Factorizing personalized Markov chains for next-basket recommendation}. In \bibinfo{booktitle}{\emph{Proceedings of the 19th International Conference on World Wide Web, {WWW} 2010, Raleigh, North Carolina, USA, April 26-30, 2010}}. \bibinfo{publisher}{{ACM}}, \bibinfo{pages}{811--820}.
\newblock
\href{https://doi.org/10.1145/1772690.1772773}{doi:\nolinkurl{10.1145/1772690.1772773}}


\bibitem[Sun et~al\mbox{.}(2019)]%
        {sun2019bert4rec}
\bibfield{author}{\bibinfo{person}{Fei Sun}, \bibinfo{person}{Jun Liu}, \bibinfo{person}{Jian Wu}, \bibinfo{person}{Changhua Pei}, \bibinfo{person}{Xiao Lin}, \bibinfo{person}{Wenwu Ou}, {and} \bibinfo{person}{Peng Jiang}.} \bibinfo{year}{2019}\natexlab{}.
\newblock \showarticletitle{BERT4Rec: Sequential Recommendation with Bidirectional Encoder Representations from Transformer}. In \bibinfo{booktitle}{\emph{Proceedings of the 28th {ACM} International Conference on Information and Knowledge Management, {CIKM} 2019, Beijing, China, November 3-7, 2019}}. \bibinfo{publisher}{{ACM}}, \bibinfo{pages}{1441--1450}.
\newblock
\href{https://doi.org/10.1145/3357384.3357895}{doi:\nolinkurl{10.1145/3357384.3357895}}


\bibitem[Tang and Wang(2018)]%
        {tang2018personalized}
\bibfield{author}{\bibinfo{person}{Jiaxi Tang} {and} \bibinfo{person}{Ke Wang}.} \bibinfo{year}{2018}\natexlab{}.
\newblock \showarticletitle{Personalized Top-N Sequential Recommendation via Convolutional Sequence Embedding}. In \bibinfo{booktitle}{\emph{Proceedings of the Eleventh {ACM} International Conference on Web Search and Data Mining, {WSDM} 2018, Marina Del Rey, CA, USA, February 5-9, 2018}}. \bibinfo{publisher}{{ACM}}, \bibinfo{pages}{565--573}.
\newblock
\href{https://doi.org/10.1145/3159652.3159656}{doi:\nolinkurl{10.1145/3159652.3159656}}


\bibitem[Tian et~al\mbox{.}(2024)]%
        {tian2024eulerformer}
\bibfield{author}{\bibinfo{person}{Zhen Tian}, \bibinfo{person}{Wayne~Xin Zhao}, \bibinfo{person}{Changwang Zhang}, \bibinfo{person}{Xin Zhao}, \bibinfo{person}{Zhongrui Ma}, {and} \bibinfo{person}{Ji{-}Rong Wen}.} \bibinfo{year}{2024}\natexlab{}.
\newblock \showarticletitle{EulerFormer: Sequential User Behavior Modeling with Complex Vector Attention}. In \bibinfo{booktitle}{\emph{Proceedings of the 47th International {ACM} {SIGIR} Conference on Research and Development in Information Retrieval, {SIGIR} 2024, Washington DC, USA, July 14-18, 2024}}. \bibinfo{publisher}{{ACM}}, \bibinfo{pages}{1619--1628}.
\newblock
\href{https://doi.org/10.1145/3626772.3657805}{doi:\nolinkurl{10.1145/3626772.3657805}}


\bibitem[Vaswani et~al\mbox{.}(2017)]%
        {vaswani2017attention}
\bibfield{author}{\bibinfo{person}{Ashish Vaswani}, \bibinfo{person}{Noam Shazeer}, \bibinfo{person}{Niki Parmar}, \bibinfo{person}{Jakob Uszkoreit}, \bibinfo{person}{Llion Jones}, \bibinfo{person}{Aidan~N. Gomez}, \bibinfo{person}{Lukasz Kaiser}, {and} \bibinfo{person}{Illia Polosukhin}.} \bibinfo{year}{2017}\natexlab{}.
\newblock \showarticletitle{Attention is All you Need}. In \bibinfo{booktitle}{\emph{Proceedings of the 30th Annual Conference on Neural Information Processing Systems, {NeurIPS} 2017, Long Beach, CA, USA, December 4-9, 2017}}. \bibinfo{pages}{5998--6008}.
\newblock
\urldef\tempurl%
\url{https://proceedings.neurips.cc/paper/2017/hash/3f5ee243547dee91fbd053c1c4a845aa-Abstract.html}
\showURL{%
\tempurl}


\bibitem[Yan et~al\mbox{.}(2019)]%
        {yan2019cosrec}
\bibfield{author}{\bibinfo{person}{An Yan}, \bibinfo{person}{Shuo Cheng}, \bibinfo{person}{Wang{-}Cheng Kang}, \bibinfo{person}{Mengting Wan}, {and} \bibinfo{person}{Julian~J. McAuley}.} \bibinfo{year}{2019}\natexlab{}.
\newblock \showarticletitle{CosRec: 2D Convolutional Neural Networks for Sequential Recommendation}. In \bibinfo{booktitle}{\emph{Proceedings of the 28th {ACM} International Conference on Information and Knowledge Management, {CIKM} 2019, Beijing, China, November 3-7, 2019}}. \bibinfo{publisher}{{ACM}}, \bibinfo{pages}{2173--2176}.
\newblock
\href{https://doi.org/10.1145/3357384.3358113}{doi:\nolinkurl{10.1145/3357384.3358113}}


\bibitem[Yue et~al\mbox{.}(2024)]%
        {yue2024linear}
\bibfield{author}{\bibinfo{person}{Zhenrui Yue}, \bibinfo{person}{Yueqi Wang}, \bibinfo{person}{Zhankui He}, \bibinfo{person}{Huimin Zeng}, \bibinfo{person}{Julian~J. McAuley}, {and} \bibinfo{person}{Dong Wang}.} \bibinfo{year}{2024}\natexlab{}.
\newblock \showarticletitle{Linear Recurrent Units for Sequential Recommendation}. In \bibinfo{booktitle}{\emph{Proceedings of the 17th {ACM} International Conference on Web Search and Data Mining, {WSDM} 2024, Merida, Mexico, March 4-8, 2024}}. \bibinfo{publisher}{{ACM}}, \bibinfo{pages}{930--938}.
\newblock
\href{https://doi.org/10.1145/3616855.3635760}{doi:\nolinkurl{10.1145/3616855.3635760}}


\bibitem[Zhai et~al\mbox{.}(2024)]%
        {zhai2024actions}
\bibfield{author}{\bibinfo{person}{Jiaqi Zhai}, \bibinfo{person}{Lucy Liao}, \bibinfo{person}{Xing Liu}, \bibinfo{person}{Yueming Wang}, \bibinfo{person}{Rui Li}, \bibinfo{person}{Xuan Cao}, \bibinfo{person}{Leon Gao}, \bibinfo{person}{Zhaojie Gong}, \bibinfo{person}{Fangda Gu}, \bibinfo{person}{Jiayuan He}, \bibinfo{person}{Yinghai Lu}, {and} \bibinfo{person}{Yu Shi}.} \bibinfo{year}{2024}\natexlab{}.
\newblock \showarticletitle{Actions Speak Louder than Words: Trillion-Parameter Sequential Transducers for Generative Recommendations}. In \bibinfo{booktitle}{\emph{Proceedings of the 41st International Conference on Machine Learning, {ICML} 2024, Vienna, Austria, July 21--27, 2024}}. \bibinfo{publisher}{OpenReview.net}.
\newblock
\urldef\tempurl%
\url{https://openreview.net/forum?id=xye7iNsgXn}
\showURL{%
\tempurl}


\end{thebibliography}

\appendix
\section{Baseline Details}
\label{baselinedetail}

\textbf{Baseline Methods.} 
We compare HyenaRec against a comprehensive set of state-of-the-art sequential recommendation models spanning recurrent-, attention-, and linear-based paradigms. All baselines are implemented under identical settings for fair comparison.

\begin{itemize}
\item \textbf{GRU4Rec:} 
An RNN-based sequential recommender using Gated Recurrent Units to capture temporal dependencies~\cite{hidasi2015session,hidasi2018recurrent}. 
It models short-term behaviors effectively.

\item \textbf{NARM:} 
A neural attentive recommendation model combining RNNs with attention to capture sequential patterns and user intents~\cite{li2017neural}. 
It integrates local and global representations for better interpretability.

\item \textbf{SASRec:} 
A self-attention-based model that adapts the Transformer architecture for sequential recommendation~\cite{kang2018self}. 
It excels at modeling long-range dependencies via causal attention, yet its quadratic complexity in sequence length limits scalability.

\item \textbf{BERT4Rec:} 
A bidirectional Transformer model trained with a cloze-style objective to utilize both left and right contexts~\cite{sun2019bert4rec}. 
This bidirectional design improves representation learning and ranking accuracy, but incurs high computational overhead during training.

\item \textbf{EulerFormer:} 
A Transformer variant that employs complex-valued attention to jointly encode semantic and positional information~\cite{tian2024eulerformer}. 
The Euler representation enhances modeling flexibility and alleviates over-smoothing issues common in standard attention mechanisms.

\item \textbf{HSTU:} 
A hierarchical sequential transducer designed for generative recommendation~\cite{zhai2024actions}. 
It captures both local and global user dynamics through stacked transducer blocks, enabling trillion-scale training with efficient autoregressive decoding.

\item \textbf{LRURec:} 
A recent linear recurrent unit model that replaces attention with diagonalizable linear recurrences~\cite{yue2024linear}. 
It achieves efficient long-context modeling with parallelizable computation and low memory cost, offering a strong balance between performance and efficiency.
\end{itemize}

\section{Implementation Details and Environments}
\label{implementation}
Key hyperparameters of HyenaRec for different datasets are summarized in Table~\ref{tab:hyperparams}. 
No sliding window truncation was applied; sequence lengths were set according to dataset characteristics.

\begin{table}[h]

\caption{HyenaRec hyperparameters for different datasets.}
\label{tab:hyperparams}
\centering
\resizebox{\linewidth}{!}{

\begin{tabular}{lcccc}
\toprule
Parameter & ML-1M & XLong & Steam & Video \\
\midrule

Max sequence length & 200 & 1000 & 50 & 50 \\
Learning rate & 1e-3 & 1e-3 & 1e-3 & 1e-3 \\
Optimizer & AdamW & AdamW & AdamW & AdamW \\
Hidden size & 64 & 64 & 64 & 64 \\
Layers & 2 & 2 & 2 & 2 \\
Dropout & 0.2 & 0.2 & 0.2 & 0.2 \\
Training batch size & 128 & 32 & 128 & 128 \\
Inference batch size & 256 & 64 & 256 & 256 \\
\bottomrule
\end{tabular}
}
\end{table}

All experiments were conducted on a machine equipped with a single RTX 3080 Ti GPU (12GB), 12 vCPU Intel(R) Xeon(R) Silver 4214R @ 2.40GHz, 90GB RAM, and CUDA version 12.2.

\section{Discussion on Other Architectures}
\label{atroushyenamodel}

In addition to the standard HyenaRec, we conducted a discussion on a variant that incorporates atrous (dilated) convolutions into the HyenaOperator. 
Specifically, the short convolution in each Operator of \textbf{HyenaVanilla}—which does not use the GLU gating or polynomial kernel—was replaced with an atrous convolution to capture multi-scale dependencies in user sequences. 
All other components of the model remained unchanged.

Table~\ref{tab:atroushyena} presents a comparison between the original HyenaVanilla and this AtrousHyena variant on several datasets. 
The results clearly show that introducing atrous convolutions leads to a slight improvement in \textbf{Recall@10} on datasets with longer sequences, such as ML-1M and XLong.

\begin{table}[h]
\caption{Comparison of HyenaVanilla and AtrousHyena. Only the short convolution in the Operator is replaced with an atrous convolution. Metrics reported are \textbf{Recall@10}.}
\label{tab:atroushyena}
\centering
% \resizebox{\linewidth}{!}{

\begin{tabular}{lcccc}
\toprule
\textbf{Model} & \textbf{ML-1M} & \textbf{XLong} & \textbf{Steam} & \textbf{Video} \\
\midrule
HyenaVanilla & 0.3162 & 0.4579 & 0.1258 & 0.0995 \\
AtrousHyena & 0.3280 & 0.4701 & 0.1326 & 0.1016 \\
\bottomrule
\end{tabular}
% }
\end{table}

Overall, replacing the short convolution in HyenaVanilla with an atrous convolution provides minor gains in \textbf{Recall@10}. 
This indicates that while multi-scale convolutional patterns can be captured, the effect on the evaluated datasets is limited. 
Alternative placements or combinations with gating and polynomial kernels may be necessary to achieve more substantial improvements.

\section{Discussion on Mamba4Rec}
\label{mamba4reccompare}
Recently, Mamba~\cite{gu2024mamba} introduced selective, data-dependent state-space operators that achieve near-linear efficiency for sequential modeling. \textbf{Mamba4Rec}~\cite{liu2024mamba4rec} adapts this idea to recommendation tasks, replacing attention or convolution modules with selective recurrence to capture long-term user behaviors efficiently.

While Mamba4Rec demonstrates competitive performance, our \textbf{HyenaRec} framework, particularly with AtrousHyena enhancements, achieves superior Recall@10 and NDCG@10 on the MovieLens-1M dataset. This improvement stems from the use of structured convolutional filters (Legendre Hyena Filters) and atrous convolutions, which provides more stable long-range modeling compared with the recurrent selective gating of Mamba4Rec.

Table~\ref{tab:mamba4rec_comparison} summarizes the comparison of representative sequential recommendation models on the MovieLens-1M and XLong datasets.

\begin{table}[h]
    \centering
    \renewcommand{\arraystretch}{1.1}
    \setlength{\tabcolsep}{4pt}
    \caption{Comparison between representative sequential recommenders on MovieLens-1M and XLong settings.}
    \label{tab:mamba4rec_comparison}
    \resizebox{\linewidth}{!}{
    \begin{tabular}{lcc}
        \toprule
        \multirow{2}{*}{\textbf{Model}} 
        & \textbf{ML-1M} 
        & \textbf{XLong} \\
        & \textbf{Recall@10 / NDCG@10} 
        & \textbf{Recall@10 / NDCG@10} \\
        \midrule
        SASRec~\cite{kang2018self} 
            & 0.3144 / 0.1829 
            & 0.4935 / 0.3095 \\
        LRURec~\cite{yue2024linear} 
            & 0.3240 / 0.1907 
            & 0.5269 / 0.3487 \\
        Mamba4Rec~\cite{liu2024mamba4rec} 
            & 0.3121 / 0.1822 
            & 0.5127 / 0.3540 \\
        HyenaRec (ours) 
            & \textbf{0.3347} / \textbf{0.2000} 
            & \textbf{0.5418} / \textbf{0.4192} \\
        \bottomrule
    \end{tabular}
    }
\end{table}

In summary, HyenaRec consistently outperforms Mamba4Rec in both accuracy and stability on long sequences, highlighting the effectiveness of our structured convolutional approach for sequential recommendation tasks.

\section{Sensitivity Analysis of Legendre Filter Order $K$}

\begin{figure}[h]
    \centering
    \includegraphics[width=\linewidth]{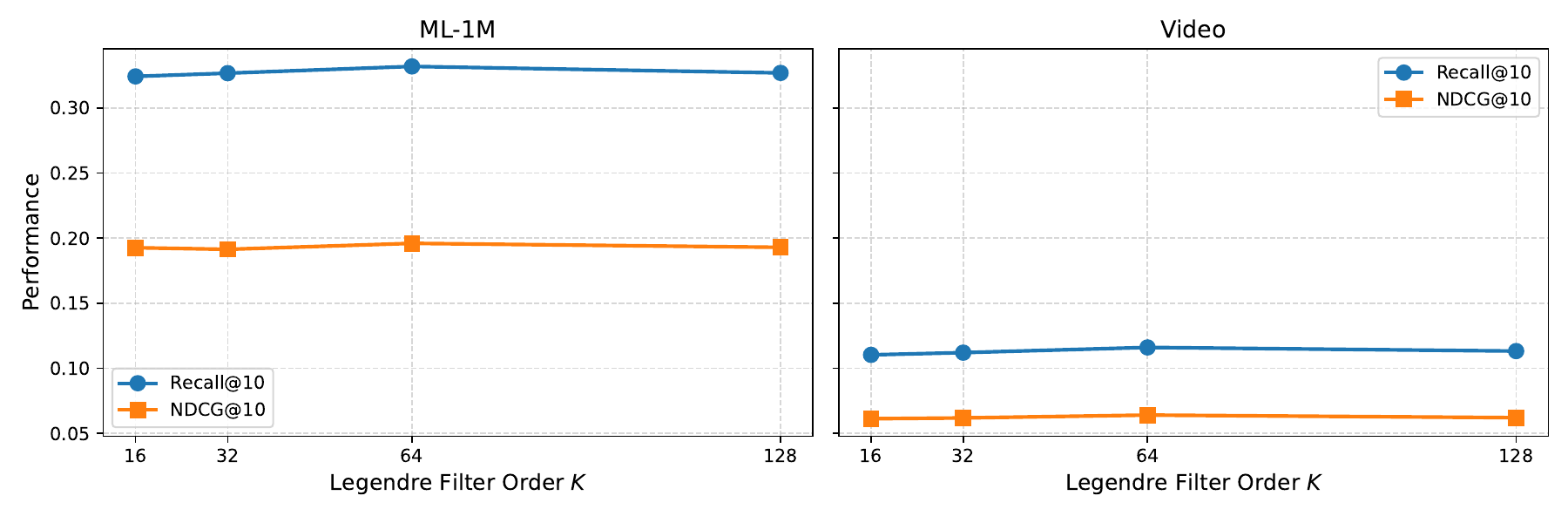}
    \caption{Performance of HyenaRec under different Legendre filter orders $K$. Left: ML-1M dataset; Right: Video dataset. Both Recall@10 and NDCG@10 are shown.}
    \label{fig:legendre_sensitivity}
\end{figure}

We conduct a sensitivity analysis of the Legendre filter order \texorpdfstring{$K$}{K} in HyenaRec to examine its influence on sequential recommendation performance. Figure~\ref{fig:legendre_sensitivity} shows the Recall@10 and NDCG@10 metrics on two datasets, ML-1M and Video, with $K$ values ranging from 16 to 128.

From Figure~\ref{fig:legendre_sensitivity}, we observe that: On ML-1M, performance increases as $K$ grows from 16 to 64, and slightly decreases at $K=128$, indicating an optimal trade-off between model expressivity and potential overfitting. On Video, a similar trend is observed, with peak performance at $K=64$, demonstrating the general effectiveness of a moderate filter order across datasets.
Overall, this analysis shows that HyenaRec is relatively robust to the choice of $K$, and that setting $K=64$ provides a good balance for both datasets.

\end{document}